\documentstyle[11pt]{article}
\begin{document}

\title{\textbf{Holographic, new agegraphic and ghost dark energy models in fractal cosmology}}

\author{K. Karami$^{1}$\thanks{E-mail: KKarami@uok.ac.ir} , Mubasher
Jamil$^{2}$\thanks{E-mail: mjamil@camp.nust.edu.pk} ,
S. Ghaffari$^{1}$, K. Fahimi$^{3}$\\\\
$^{1}$\small{Department of Physics, University of Kurdistan,
Pasdaran St., Sanandaj, Iran}\\$^{2}$\small{Center for Advanced
Mathematics and Physics (CAMP), National University} \\\small{of
Sciences and Technology (NUST), Islamabad,
Pakistan}\\$^{3}$\small{Department of Physics, Sanandaj Branch,
Islamic Azad University, Sanandaj, Iran}}

\maketitle

\begin{abstract}
We investigate the holographic, new agegraphic and ghost dark energy
models in the framework of fractal cosmology. We consider a fractal
FRW universe filled with the dark energy and dark matter. We obtain
the equation of state parameters of the selected dark energy models
in the ultraviolet regime and discuss on their implications.
\end{abstract}

\section{Fractal cosmology}
Recently, Calcagni \cite{Calcagni1,Calcagni2} gave a quantum gravity
in a fractal universe and then investigated cosmology in that
framework. That theory is Lorentz invariant, power-counting
renormalizable and free from ultraviolet divergence. In the present
paper, we study few DE models (including holographic, new agegraphic
and ghost DE) in the framework of fractal cosmology proposed by
Calcagni. We calculate their equation of state parameters and
discuss their physical implications via graphs and conclude in the
end.

The action of Einstein gravity in a fractal spacetime is given by
\cite{Calcagni1,Calcagni2}
\begin{equation}\label{F-action}
S=\int{\rm
d}\varrho(x)\sqrt{-g}\left(\frac{R-2\Lambda-\omega\partial_{\mu}v\partial^{\mu}v}{2\kappa^2}+\mathcal{L}_{\rm
m}\right),
\end{equation}
where $\kappa^2=8\pi G$. Also $G$, $g$, $R$, $\Lambda$ and
$\mathcal{L}_{\rm m}$ are the gravitational constant, determinant of
metric $g_{\mu\nu}$, Ricci scalar, cosmological constant and
Lagrangian density of the total matter inside the universe,
respectively. The quantities $v$ and $\omega$ are the fractal
function and fractal parameter, respectively. Note that ${\rm
d}\varrho(x)$ is Lebesgue-Stieltjes measure generalizing the
standard 4-dimensional measure ${\rm d}^4x$. The scaling dimension
of $\varrho$ is $[\varrho]=-4\alpha$, where $\alpha>0$ is a
parameter. The values of $\alpha$ in the infrared (IR) and
ultraviolet (UV) regimes are $\alpha_{\rm IR}=1$ and $\alpha_{\rm
UV}=1/2$, respectively.

Taking the variation of the action (\ref{F-action}) with respect to
the Friedmann-Robertson-Walker (FRW) metric $g_{\mu\nu}$, one can
obtain the Friedmann equations in a fractal universe as
\cite{Calcagni2}
\begin{equation}\label{Fr01-F}
H^2+\frac{k}{a^2}+H\frac{\dot{v}}{v}-\frac{\omega}{6}\dot{v}^2=\frac{1}{3M_p^2}\rho+\frac{\Lambda}{3},
\end{equation}
\begin{equation}\label{Fr02-F}
\dot{H}+H^2-H\frac{\dot{v}}{v}+\frac{\omega}{3}\dot{v}^2-\frac{1}{2}\frac{\Box
v}{v}=-\frac{1}{6M_p^2}(\rho+3p)+\frac{\Lambda}{3},
\end{equation}
where $H=\dot{a}/a$ is the Hubble parameter and $M_p$ is the reduced
Planck mass $M_p^{-2}=8\pi G$. Also $\rho$ and $p$ are the total
energy density and pressure of the perfect fluid inside the
universe. The parameter $k=0,1,-1$ represent a flat, closed and open
FRW universe, respectively. Note that when $v=1$, Eqs.
(\ref{Fr01-F}) and (\ref{Fr02-F}) transform to the standard
Friedmann equations in Einstein GR.

The continuity equation in fractal cosmology takes the form
\cite{Calcagni2}
\begin{equation}\label{Conteq}
\dot{\rho}+\left(3H+\frac{\dot{v}}{v}\right)(\rho+p)=0.
\end{equation}
The gravitational constraint in a fractal universe is given by
\cite{Calcagni2}
\begin{equation}\label{Graveq11}
\dot{H}+3H^2+\frac{2k}{a^2}+\frac{\Box
v}{v}+H\frac{\dot{v}}{v}+\omega(v\Box v-\dot{v}^2)=0.
\end{equation}
Taking a timelike fractal profile $v=t^{-\beta}$ \cite{Calcagni2},
where $\beta=4(1-\alpha)$ is the fractal dimension, for a flat FRW
metric ($k=0$) one can obtain
\begin{equation}
\frac{\Box
v}{v}=\frac{\beta}{t}\left(3H-\frac{1+\beta}{t}\right).\label{dalv}
\end{equation}
The values of $\beta$ in the IR and UV regimes are $\beta_{\rm
IR}=0$ and $\beta_{\rm UV}=2$, respectively.

Using $v=t^{-\beta}$ and Eq. (\ref{dalv}), the Friedmann equations
(\ref{Fr01-F}) and (\ref{Fr02-F}) for a flat FRW universe reduce to
\begin{equation}\label{Friedeq01}
H^2-\frac{\beta}{t}H-\frac{\omega\beta^2}{6t^{2(\beta+1)}}=\frac{1}{3M_p^2}\rho+\frac{\Lambda}{3},
\end{equation}
\begin{equation}\label{Friedeq02}
\dot{H}+H^2-\frac{\beta}{2t}H+\frac{\beta(\beta+1)}{2t^2}
+\frac{\omega\beta^2}{3t^{2(\beta+1)}}=-\frac{1}{6M_p^2}(\rho+3p)+\frac{\Lambda}{3}.
\end{equation}
For $v=t^{-\beta}$, the continuity equation (\ref{Conteq}) takes the
form
\begin{equation}\label{Conteq1}
\dot{\rho}+\left(3H-\frac{\beta}{t}\right)(\rho+p)=0,
\end{equation}
and the gravitational constraint (\ref{Graveq11}) in a fractal flat
FRW universe with the help of Eq. (\ref{dalv}) yields
\begin{equation}\label{Graveq1}
\dot{H}+3H^2+\left(2+\frac{3\omega}{t^{2\beta}}\right)\frac{\beta}{t}H-
\frac{\beta(\beta+1)}{t^2}-\frac{\omega\beta(2\beta+1)}{t^{2(\beta+1)}}=0.
\end{equation}
Note that in the IR regime $(\beta=0)$, Eqs. (\ref{Friedeq01}) to
(\ref{Conteq1}) recover the corresponding equations in standard
Einstein GR (no gravitational constraint).

\begin{figure}
\includegraphics{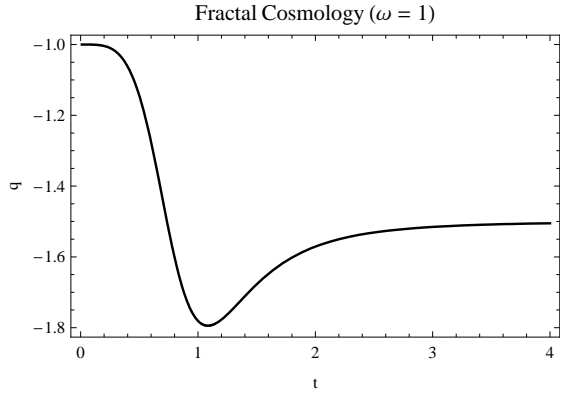}
      \vspace{5cm}
\caption[]{Time evolution of the deceleration parameter in fractal
cosmology, Eq. (\ref{qw}), for $\omega=1$.}
         \label{q-w_1}
   \end{figure}

\begin{figure}
\includegraphics{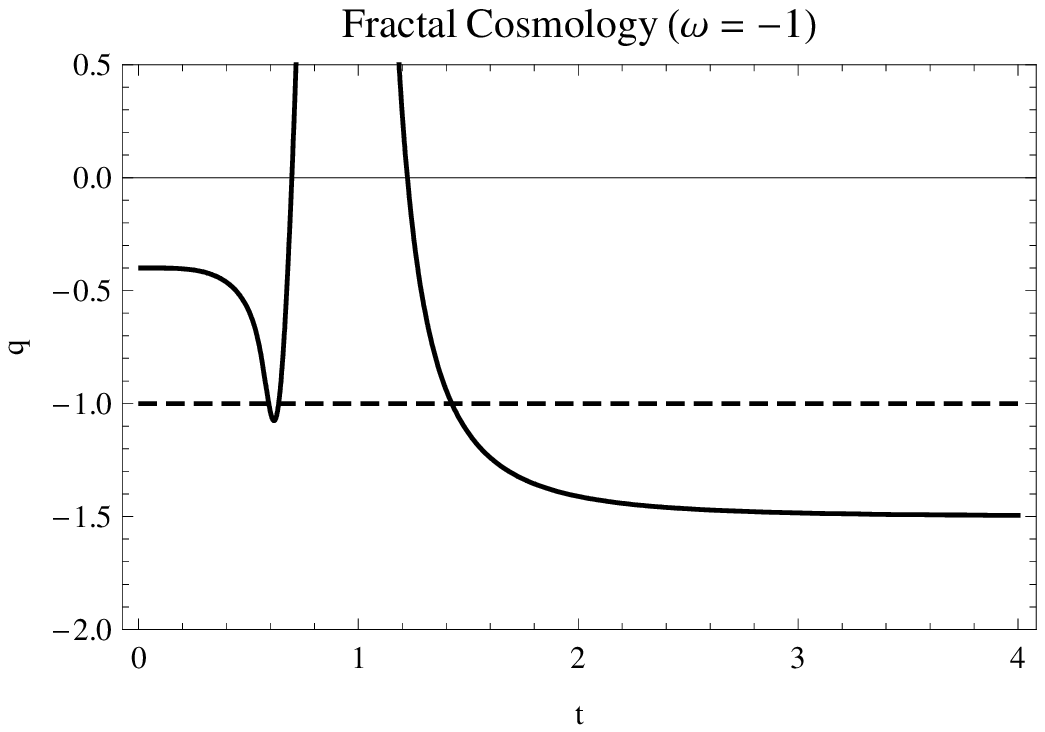}
      \vspace{5cm}
\caption[]{Same as Fig. \ref{q-w_1}, for $\omega=-1$.}
         \label{q-w_-1}
   \end{figure}

In the UV regime $(\beta=2)$ \cite{Calcagni2} with no cosmological
constant $(\Lambda=0)$, the Friedmann equations (\ref{Friedeq01})
and (\ref{Friedeq02}) yield
\begin{equation}\label{Friedeq03}
H^2-\frac{2}{t}H-\frac{2\omega}{3t^6}=\frac{1}{3M_p^2}\rho,
\end{equation}
\begin{equation}\label{Friedeq04}
\dot{H}+H^2-\frac{H}{t}+\frac{3}{t^2}+\frac{4\omega}{3t^6}=-\frac{1}{6M_p^2}(\rho+3p).
\end{equation}
We consider a universe containing the DE density $\rho_D$ and the
pressureless dark matter (DM), $p_{\rm m}=0$. Using Eq.
(\ref{Conteq1}), the energy equations for the DE and DM in the UV
regime $(\beta=2)$ reduce to
\begin{equation}
\dot{\rho}_{D}+\left(3H-\frac{2}{t}\right)(1+w_{D})\rho_{D}=0,\label{eqDE}
\end{equation}
\begin{equation}
\dot{\rho}_{\rm m}+\left(3H-\frac{2}{t}\right)\rho_{\rm
m}=0,\label{eqCDM}
\end{equation}
 where $w_D=p_D/\rho_D$ is the equation of state (EoS) parameter of the DE.
For the DM from Eq. (\ref{eqCDM}) one can get $\rho_{\rm
m}=\rho_{\rm m_0}t^2/a^3$, where $\rho_{\rm m_0}$ is the present
value of the DM energy density.
 The gravitational constraint (\ref{Graveq1}) in
the UV regime $(\beta=2)$ yields
\begin{equation}
\dot{H}+3H^2+\left(2+\frac{3\omega}{t^4}\right)\frac{2}{t}H-
\frac{6}{t^2}-\frac{10\omega}{t^6}=0.\label{Graveq2}
\end{equation}
Solving the above differential equation gives the Hubble parameter
as \cite{Calcagni2}
\begin{equation}\label{Hw}
H(t)=-\frac{2}{t}-\frac{22\omega}{13t^5} \frac{{_1}{\rm
F}_1\Big(\frac{15}{4};\frac{17}{4};\frac{3\omega}{2t^4}\Big)}{{_1}{\rm
F}_1\Big(\frac{11}{4};\frac{13}{4};\frac{3\omega}{2t^4}\Big)},
\end{equation}
and the scale factor is obtained as
\begin{equation}
a(t)=\frac{1}{t^2}~{_1}{\rm
F}_1\left(\frac{11}{4};\frac{13}{4};\frac{3\omega}{2t^4}\right)^{1/3},\label{aw}
\end{equation}
where ${_1}{\rm F}_1$ is the hypergeometric function of the first
kind. The deceleration parameter is obtained as
\begin{equation}\label{qw}
q=-1-\frac{\dot{H}}{H^2}=-\frac{{\rm I}_q}{{\rm II}_q},
\end{equation}
where
\begin{eqnarray}
{\rm I}_q&=&169 (3 t^4 + \omega) (t^4 + 2\omega) {_1}{\rm
F}_1\left(\frac {11} {4};\frac {13} {4};\frac {3\omega} {2
t^4}\right)^2\nonumber\\&&+52\omega (2 t^4 + \omega) {_1}{\rm
F}_1\left(\frac {11} {4};\frac {13} {4};\frac {3\omega} {2
t^4}\right){_1}{\rm F}_1\left(\frac {11} {4};\frac {17}
{4};\frac{3\omega} {2 t^4}\right)\nonumber\\&&-16\omega^2~{_1}{\rm
F}_1\left(\frac {11}{4};\frac {17} {4};\frac {3\omega}
{2t^4}\right)^2,
\end{eqnarray}
\begin{equation}
{\rm II}_q=2 \left[13 t^4~ {_1}{\rm F}_1\left(\frac {11}
{4};\frac{13}{4};\frac {3\omega}{2t^4}\right)+11\omega~{_1}{\rm
F}_1\left(\frac {15} {4};\frac {17} {4};\frac {3\omega} {2
t^4}\right)\right]^2.
\end{equation}
At early times ($t\rightarrow 0$), from Eq. (\ref{qw}) the
deceleration parameter behaves like
\begin{equation}
q\sim-1-\frac{5t^4}{2\omega},
\end{equation}
for $\omega>0$ and
\begin{equation}
q\sim-\frac{2}{5},
\end{equation}
for $\omega<0$. At late times ($t\rightarrow +\infty$), the
deceleration parameter for the both $\omega>0$ and $\omega<0$ yields
\begin{equation}
q\sim-\frac{3}{2}.
\end{equation}
Time evolution of the deceleration parameter (\ref{qw}) is plotted
in Figs. \ref{q-w_1} and \ref{q-w_-1} for $\omega=+1$ and
$\omega=-1$, respectively. Figure \ref{q-w_1} shows that for
$\omega=+1$ the model describes an expanding super-accelerating
universe. At $t=0$, the expansion is de Sitter, i.e. $q=-1$. Figure
\ref{q-w_-1} reveal that the model with $\omega=-1$ is a universe
which expands in acceleration. At some points ($t=0.59,0.64,1.42$)
the expansion is de Sitter ($q=-1$). Also at $t=0.70$ and $1.22$ we
have the opposite (cosmic acceleration to deceleration, i.e.
$q<0\rightarrow q>0$) and direct ($q>0\rightarrow q<0$) transitions,
respectively.
\section{Holographic DE in a fractal universe}
\begin{figure}
\includegraphics{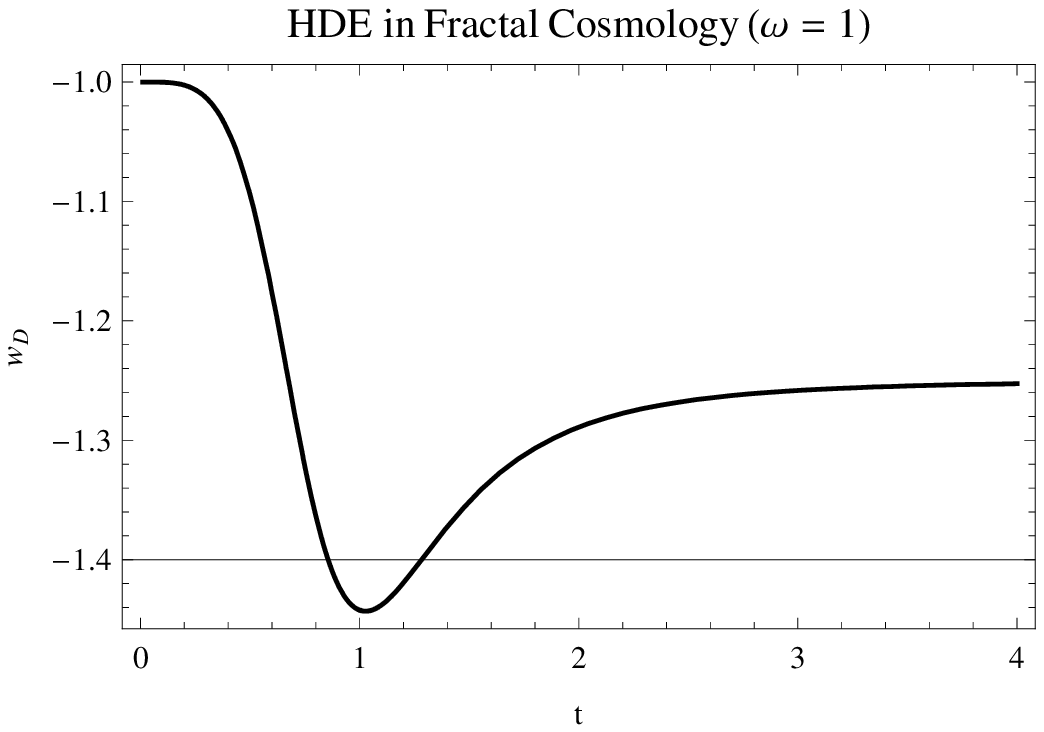}
      \vspace{5cm}
\caption[]{Time evolution of the EoS parameter of the HDE in fractal
cosmology, Eq. (\ref{omegaHDE}), for $\omega=1$.}
         \label{HDE-wD-t-w_1}
   \end{figure}

\begin{figure}
\includegraphics{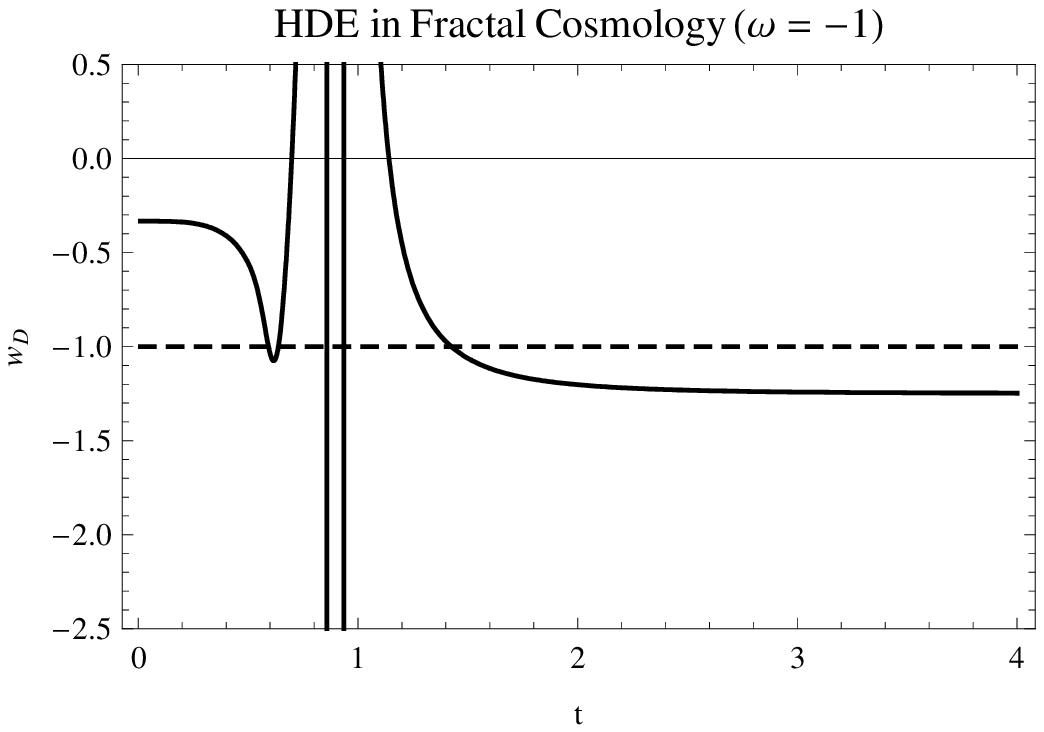}
      \vspace{5cm}
\caption[]{Same as Fig. \ref{HDE-wD-t-w_1}, for $\omega=-1$.}
         \label{HDE-wD-t-w_-1}
   \end{figure}

The holographic DE (HDE) density is given by \cite{Li}
\begin{equation}\label{HDE}
\rho_D=3c^2M_p^2L^{-2},
\end{equation}
where $L$ is the IR-cutoff of the universe. For the Hubble horizon,
i.e. $L=H^{-1}$, taking time derivative of both sides of Eq.
(\ref{HDE}) yields
\begin{equation}
\frac{\dot{\rho}_D}{\rho_D}=2\frac{\dot{H}}{H}.
\end{equation}
Substituting the above relation into Eq. (\ref{eqDE}) and using
(\ref{Hw}) one can get the EoS parameter of the HDE as
\begin{equation}\label{omegaHDE}
w_D=-1-\frac{\dot{H}}{H}\left(\frac{2t}{3tH-2}\right)=\frac{{\rm
I}_{\rm HDE}}{{\rm II}_{\rm HDE}},
\end{equation}
where
\begin{eqnarray}
{\rm I}_{\rm HDE}&=&-169(5t^8+10t^4\omega+3\omega^2){_1}{\rm
F}_1\left(\frac {11} {4};\frac{13}{4};\frac
{3\omega}{2t^4}\right)^2\nonumber\\&&-26t^4\omega~{_1}{\rm F}_1\left
(\frac {11} {4};\frac{13}{4};\frac {3\omega}{2t^4}\right){_1}{\rm
F}_1\left(\frac {11} {4};\frac{17}{4};\frac
{3\omega}{2t^4}\right)\nonumber\\&&+12\omega^2~{_1}{\rm F}_1\left
(\frac {11} {4};\frac{17}{4};\frac {3\omega}{2t^4}\right)^2,
\end{eqnarray}
\begin{eqnarray}
{\rm II}_{\rm HDE}&=&\left[13t^4~{_1}{\rm F}_1\left(\frac {11}
{4};\frac{13}{4};\frac {3\omega}{2t^4}\right)+11\omega~{_1}{\rm
F}_1\left(\frac {15}{4};\frac{17}{4};\frac
{3\omega}{2t^4}\right)\right]\nonumber\\&\times&\left[52t^4~{_1}{\rm
F}_1\left(\frac {11}{4};\frac{13}{4};\frac
{3\omega}{2t^4}\right)+33\omega~{_1}{\rm F}_1\left(\frac
{15}{4};\frac{17}{4};\frac {3\omega}{2t^4}\right)\right].
\end{eqnarray}
Time evolution of the EoS parameter (\ref{omegaHDE}) for the cases
$\omega=+1$ and $\omega=-1$ are plotted in Figs. \ref{HDE-wD-t-w_1}
and \ref{HDE-wD-t-w_-1}, respectively. These figures show that at
late times ($t\rightarrow +\infty$), the EoS parameter of the HDE
for the both $\omega=+1$ and $\omega=-1$ yields $w_D\sim-1.25$.
Figure \ref{HDE-wD-t-w_1} shows that $w_D$ at early times
($t\rightarrow 0$) behaves like cosmological constant, i.e.
$w_D\sim-1$, and after that behaves as phantom type DE, i.e.
$\omega_D<-1$. Figure \ref{HDE-wD-t-w_-1} clears that $w_D$ at early
and late times behaves like quintessence ($w_D>-1$) and phantom
($w_D<-1$) models of DE, respectively. Figure \ref{HDE-wD-t-w_-1} in
more detail illustrates that at $t=0.59$, $0.64$ and $1.42$ we have
three transitions containing (quintessence~$\rightarrow$~phantom),
(phantom~$\rightarrow$~quintessence) and
(quintessence~$\rightarrow$~phantom), respectively. Furthermore, at
$t=(0.70,1.14)$, the EoS parameter behaves like the pressureless
dust (or dark) matter, i.e. $w_D=0$, which is a wrong EoS parameter
of DE. This result has been already obtained by Hsu \cite{Hsu} for
the HDE model with the IR cut-off $L=H^{-1}$ in standard cosmology.
\section{New agegraphic DE in a fractal universe}
\begin{figure}
\includegraphics{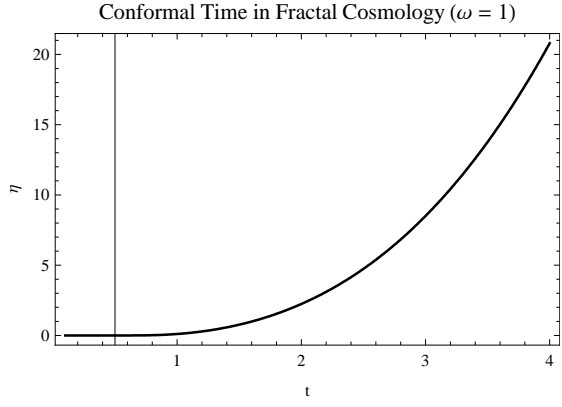}
      \vspace{5cm}
\caption[]{Time evolution of the conformal time $\eta$ in fractal
cosmology, Eq. (\ref{Etaeq}), for $\omega=1$.}
         \label{NADE-eta-t-w_1}
   \end{figure}

\begin{figure}
\includegraphics{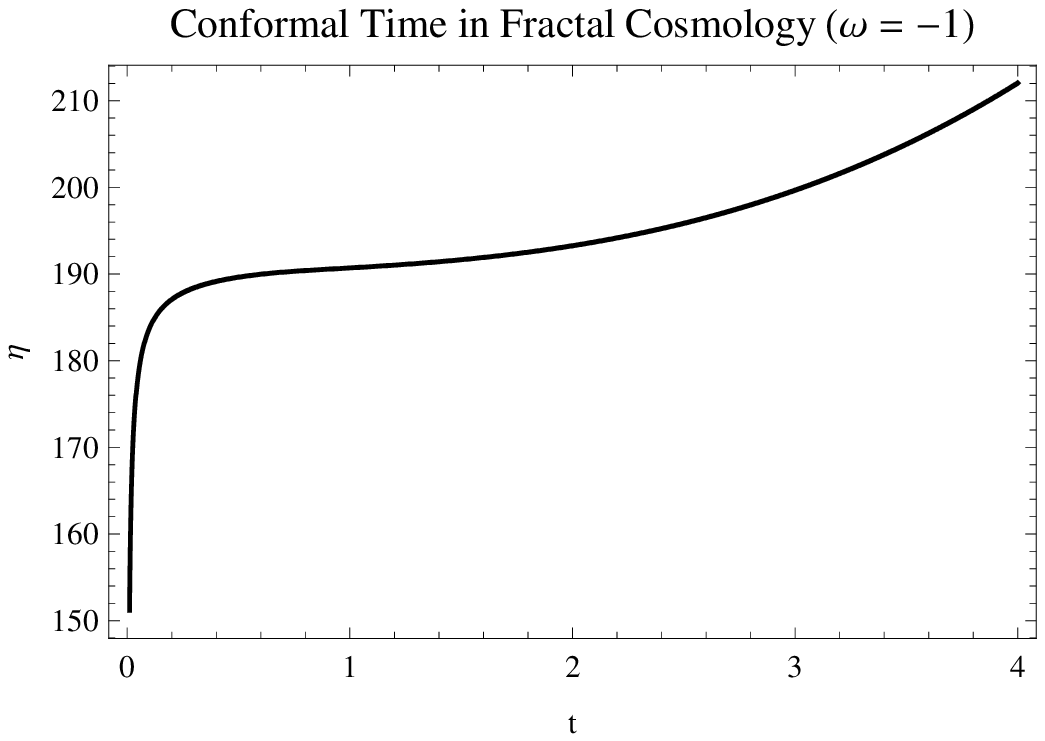}
      \vspace{5cm}
\caption[]{Same as Fig. \ref{NADE-eta-t-w_1}, for $\omega=-1$.}
         \label{NADE-eta-t-w_-1}
   \end{figure}

The new agegraphic DE (NADE) density is given by \cite{Wei1}
\begin{equation}\label{NADE}
\rho_D=3n^2M_p^2\eta^{-2},
\end{equation}
where $\eta$ is a conformal time defined as
\begin{equation}\label{Etaeq}
\eta=\int_{0}^t\frac{{\rm d}t}{a}=\int_{0}^a\frac{{\rm d}a}{Ha^2}.
\end{equation}
Time evolution of the conformal time $\eta$ for the cases
$\omega=+1$ and $\omega=-1$ are plotted in Figs.
\ref{NADE-eta-t-w_1} and \ref{NADE-eta-t-w_-1}, respectively. The
both figures show that the conformal time $\eta$ increases with
increasing the time.

\begin{figure}
\includegraphics{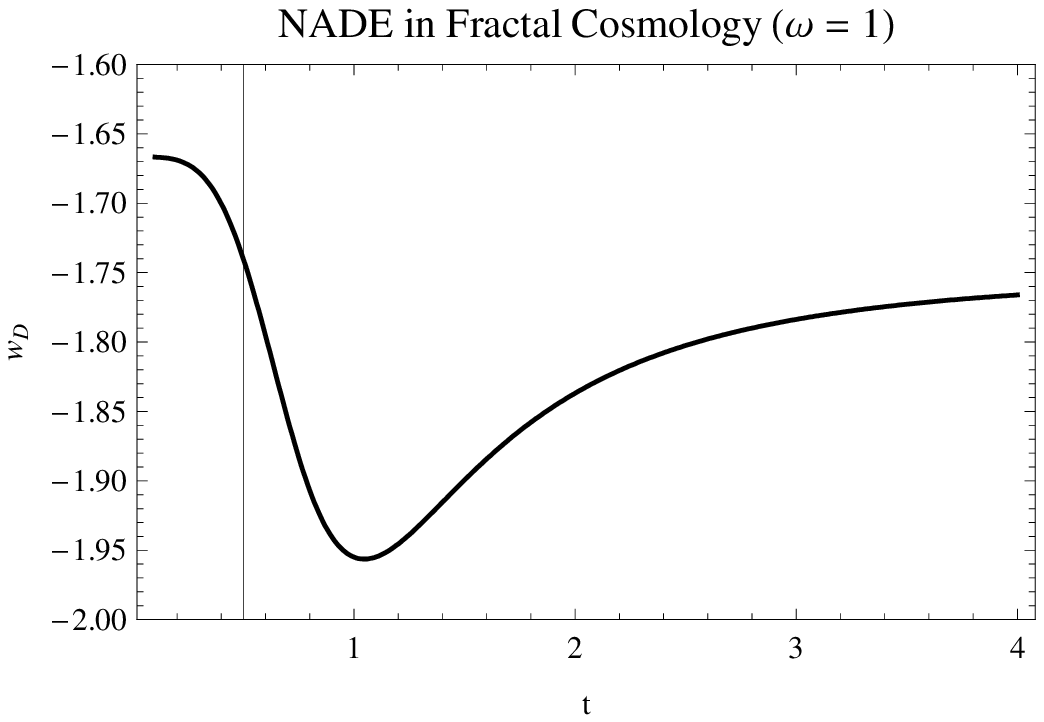}
      \vspace{5cm}
\caption[]{Time evolution of the EoS parameter of the NADE in
fractal cosmology, Eq. (\ref{omegaNADE}), for $\omega=1$.}
         \label{NADE-wD-t-w_1}
   \end{figure}

\begin{figure}
\includegraphics{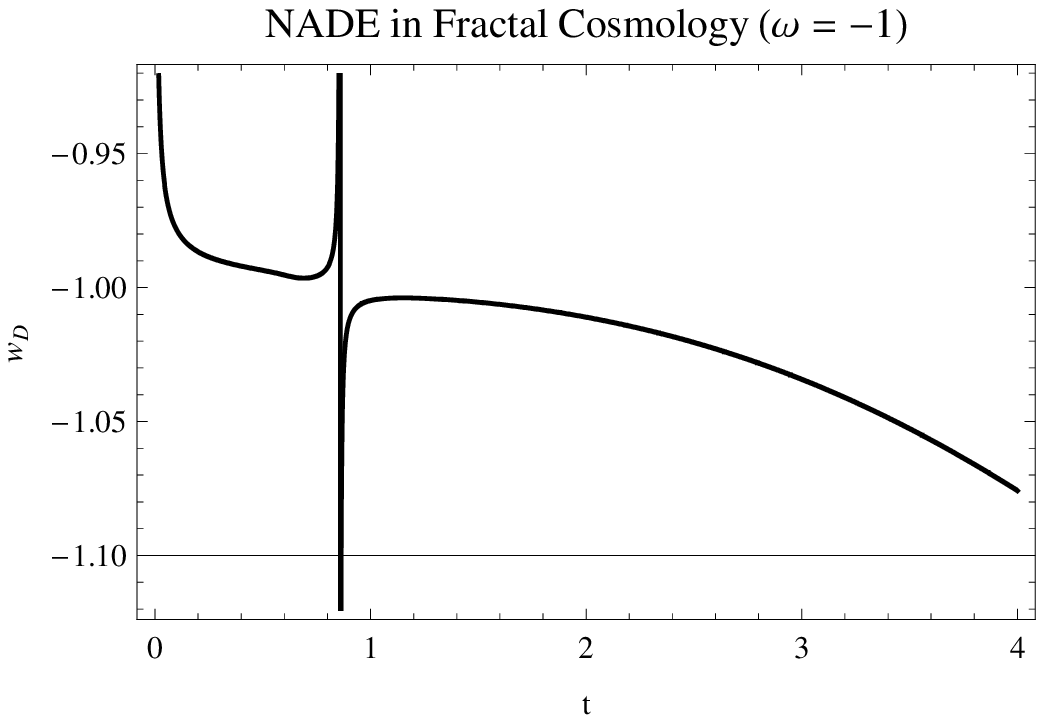}
      \vspace{5cm}
\caption[]{Same as Fig. \ref{NADE-wD-t-w_1}, for $\omega=-1$.}
         \label{NADE-wD-t-w_-1}
   \end{figure}

Taking time derivative of Eq. (\ref{NADE}) and using
$\dot{\eta}=1/a$ gives
\begin{equation}
\frac{\dot{\rho}_D}{\rho_D}=-\frac{2}{a\eta}.\label{rhodotNADE}
\end{equation}
Inserting Eq. (\ref{rhodotNADE}) into (\ref{eqDE}), one can obtain
the EoS parameter of the NADE as
\begin{equation}\label{omegaNADE}
w_D=-1+\frac{1}{a\eta}\left(\frac{2t}{3tH-2}\right).
\end{equation}
Time evolution of the EoS parameter (\ref{omegaNADE}) for the cases
$\omega=+1$ and $\omega=-1$ are plotted in Figs. \ref{NADE-wD-t-w_1}
and \ref{NADE-wD-t-w_-1}, respectively. Figure \ref{NADE-wD-t-w_1}
shows that $w_D$ always behaves like phantom type DE and at late
times ($t\rightarrow +\infty$) we have $w_D\sim-1.75$. Figure
\ref{NADE-wD-t-w_-1} clears that $w_D$ for $t<0.86$ and $t>0.86$
behaves like quintessence ($w_D>-1$) and phantom ($w_D<-1$) types of
DE, respectively. At $t=0.86$ we have $w_D=-1.52$ which lies beyond
the frame of Fig. \ref{NADE-wD-t-w_-1}.
\section{Ghost DE in a fractal universe}
\begin{figure}
\includegraphics{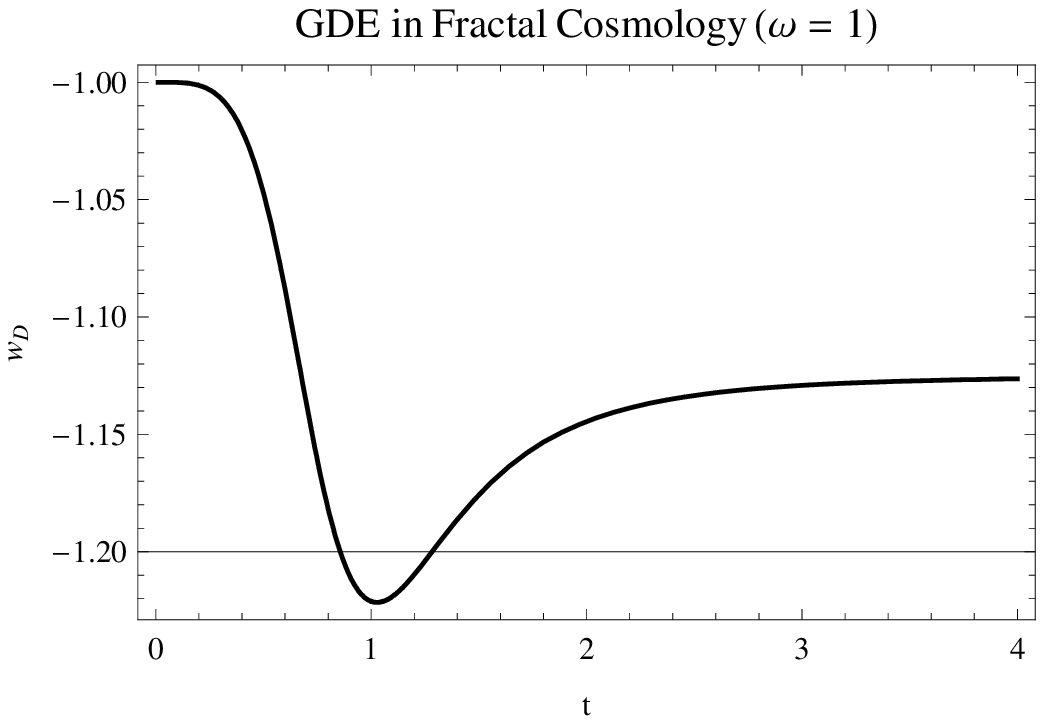}
      \vspace{5cm}
\caption[]{Time evolution of the EoS parameter of the GDE in fractal
cosmology, Eq. (\ref{omegaGDE}), for $\omega=1$.}
         \label{GDE-wD-t-w_1}
   \end{figure}

\begin{figure}
\includegraphics{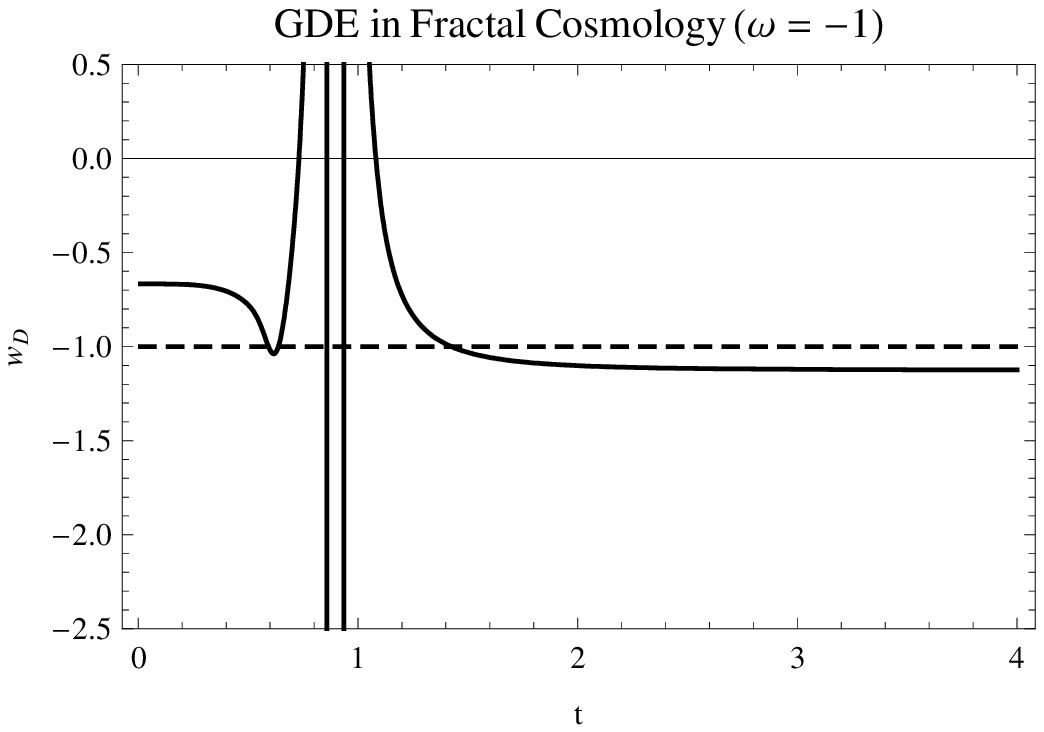}
      \vspace{5cm}
\caption[]{Same as Fig. \ref{GDE-wD-t-w_1}, for $\omega=-1$.}
         \label{GDE-wD-t-w_-1}
   \end{figure}

The ghost DE (GDE) density is given by \cite{Cai}
\begin{equation}\label{GDE}
\rho_D=\alpha H,
\end{equation}
where $\alpha$ is a constant. Taking time derivative of Eq.
(\ref{GDE}) yields
\begin{equation}
\frac{\dot{\rho}_D}{\rho_D}=\frac{\dot{H}}{H}.
\end{equation}
Substituting this into Eq. (\ref{eqDE}) and using (\ref{Hw}) yields
the EoS parameter of the GDE as
\begin{equation}\label{omegaGDE}
w_D=-1-\frac{\dot{H}}{H}\left(\frac{t}{3tH-2}\right)=\frac{{\rm
I}_{\rm GDE}}{{\rm II}_{\rm GDE}},
\end{equation}
where
\begin{eqnarray}
{\rm I}_{\rm GDE}=13~{_1}{\rm F}_1\left(\frac {11}
{4};\frac{13}{4};\frac
{3\omega}{2t^4}\right)\left[-13(9t^8+17t^4\omega+6\omega^2){_1}{\rm
F}_1\left(\frac {11} {4};\frac{13}{4};\frac
{3\omega}{2t^4}\right)\right.~\nonumber\\\left.+12\omega(t^4+\omega)~{_1}{\rm
F}_1\left(\frac {11} {4};\frac{17}{4};\frac
{3\omega}{2t^4}\right)\right],
\end{eqnarray}
\begin{equation}
{\rm II}_{\rm GDE}=2~{\rm II}_{\rm HDE}.
\end{equation}
Time evolution of the EoS parameter (\ref{omegaGDE}) for the cases
$\omega=+1$ and $\omega=-1$ are plotted in Figs. \ref{GDE-wD-t-w_1}
and \ref{GDE-wD-t-w_-1}, respectively. Figures \ref{GDE-wD-t-w_1}
and \ref{GDE-wD-t-w_-1} show that the EoS parameter of the GDE
behaves like the HDE model (see Figs. \ref{HDE-wD-t-w_1} and
\ref{HDE-wD-t-w_-1}). At late times ($t\rightarrow +\infty$), the
EoS parameter of the GDE for the both $\omega=+1$ and $\omega=-1$
yields $w_D\sim-1.125$.
\section{Conclusions}
Here, we investigated the holographic, new agegraphic and ghost DE
models in the framework of fractal cosmology. We calculated the
equation of state parameters of these models and plotted them in
graphs. We performed our analysis under the limits $\Lambda=0$ and
$\beta=2$ (UV regime). We got acceleration and super-acceleration
phases of expansion in early and late time evolution. We obtained
transition phases from acceleration to deceleration and backwards.
At different times, we got phantom and quintessence types of DE.

\clearpage
\clearpage
\clearpage
\clearpage
\clearpage

\end{document}